\documentclass[journal=nalefd,manuscript=article]{achemso}

\usepackage{graphicx}
\usepackage{siunitx}

\author{G\"u{}nter Kewes}
\email{gkewes@physik.hu-berlin.de}
\affiliation[Humboldt-Universit\"at zu Berlin, AG Nanooptik]
{Humboldt-Universit\"at zu Berlin, Institut f\"ur Physik, AG Nanooptik, Newtonstr. 15, D-12489 Berlin, Germany}
\author{Rogelio Rodr\'{i}guez-Oliveros}
\affiliation[Humboldt-Universit\"at zu Berlin, AG Theoretische Optik \& Photonik]
{Humboldt-Universit\"at zu Berlin, Institut f\"ur Physik, AG Theoretische Optik \& Photonik, Newtonstr. 15, D-12489 Berlin, Germany}
\author{Kathrin H\"o{}fner}
\affiliation[Humboldt-Universit\"at zu Berlin, AG Theoretische Optik \& Photonik]
{Humboldt-Universit\"at zu Berlin, Institut f\"ur Physik, AG Theoretische Optik \& Photonik, Newtonstr. 15, D-12489 Berlin, Germany}
\author{Alexander Kuhlicke}
\affiliation[Humboldt-Universit\"at zu Berlin, AG Nanooptik]
{Humboldt-Universit\"at zu Berlin, Institut f\"ur Physik, AG Nanooptik, Newtonstr. 15, D-12489 Berlin, Germany}
\author{Oliver Benson}
\affiliation[Humboldt-Universit\"at zu Berlin, AG Nanooptik]
{Humboldt-Universit\"at zu Berlin, Institut f\"ur Physik, AG Nanooptik, Newtonstr. 15, D-12489 Berlin, Germany}
\author{Kurt Busch}
\affiliation[Humboldt-Universit\"at zu Berlin, AG Theoretische Optik \& Photonik]
{Humboldt-Universit\"at zu Berlin, Institut f\"ur Physik, AG Theoretische Optik \& Photonik, Newtonstr. 15, D-12489 Berlin, Germany}
\alsoaffiliation{Max-Born Institut, Max-Born-Strasse 2a, D-12489 Berlin, Germany }

\title{A fully nanoscopic dielectric laser}

\begin{document}

\begin{abstract}
In this article, we introduce the concept of a dielectric nanolaser that is nanoscopic in all spatial dimensions. Our proposal is based on dielectric nanoparticles of high refractive index, e.g., silicon, acting as a (passive) cavity (without intrinsic gain) that is decorated with a thin film of organic gain media. Its resonance frequencies can be tuned over the entire visible range and bright and dark modes can be addressed. So called ''magnetic'' modes can be utilized, which makes this dielectric nanolaser a complementary source of coherent nearfields similar to the surface plasmon laser (which exploits electric modes). 
The small intrinsic losses in silicon yield relatively high quality factors and low non-radiative decay rates of emitters close to the cavity, both of which will lead to low thresholds. As we show in this work, the dielectric nanolaser exhibits certain advantages relative to nanowire lasers and spasers, such as reduced laser threshold, short switch-on times, size and design flexibility. The dielectric nanolaser is compatible with standard lithographic fabrication approaches and its relative simple design may allow for easy testing and realization of the concept. Thus, the silicon nanolaser might soon find many applications in nanooptics and metamaterials.  
\end{abstract}


In order to achieve high integration densities for optoelectronic structures, research aims for lasers to become smaller and smaller. For fully dielectric lasers a natural limit appeared to be found, when semiconductor nanowire lasers were demonstrated \cite{Yan2009,Ma2013,Saxena2013}. The concept of surface plasmon lasers (spasers) overcame this size limitation by utilizing plasmons instead of photons \cite{Bergman2003}. While the potential relevance of the spaser concept is out of question, as it could enable a plethora of applications in nanooptics \cite{Stockman2010,Stockman2011a} and metamaterials \cite{Hess2012b}, spasers suffer from high spasing/lasing thresholds that could only be reduced when demanding fabrication requirements were employed \cite{Lu2012a}. Furthermore, only few publications \cite{Noginov2009,Meng2013a} address the possibly most attractive design of fully nanoscopic spaser structures that exploit localized plasmon modes in nanoparticles. Recently, it was even questioned, that the spaser can provide any benefit with respect to other small lasers due to their high thresholds \cite{Khurgin2014}.

As we pointed out in our recently published theory for nanoscopic spasers \cite{Kewes2014a}, which is based on fully electrodynamic (and analytic) Mie-theory, there are ways to significantly improve the performance of spasers. However, the question arises, if there exist other and potentially more feasible alternatives. Since our above theory allows not only the treatment of metallic spherical plasmon-cavities but, in principle, any sphere supporting Mie-modes, we investigate, in this work, spherical silicon nanoparticles as cavities. Specifically, we will show through a comparative study of gold, silver and silicon cavities and their associated lasing performance, that certain aspects of dielectric nanolasers based on silicon exhibit superior performance characteristics.  

The heart of our nanolaser proposal is an improved resonator. Being well known for microwave antenna designs, dielectric spheres have only recently been introduced as useful nanoantennas \cite{Kuznetsov2012,Coenen2013}. In fact, silicon nanospheres support ''magnetic'' (TE) and ''electric'' (TM) Mie-modes over the entire visible spectral range when the diameter is tuned from 100-\SI{200}{nm}. As we will show, a very thin layer of around \SI{1}{nm} of gain-medium, possibly deposited by the well-established layer-by-layer approach \cite{DeLacy2013}, can provide sufficient gain for such a cavity to turn into a fully nanoscopic laser. 
 
\begin{figure}[h]
 \centering
\includegraphics[width=0.85\textwidth]{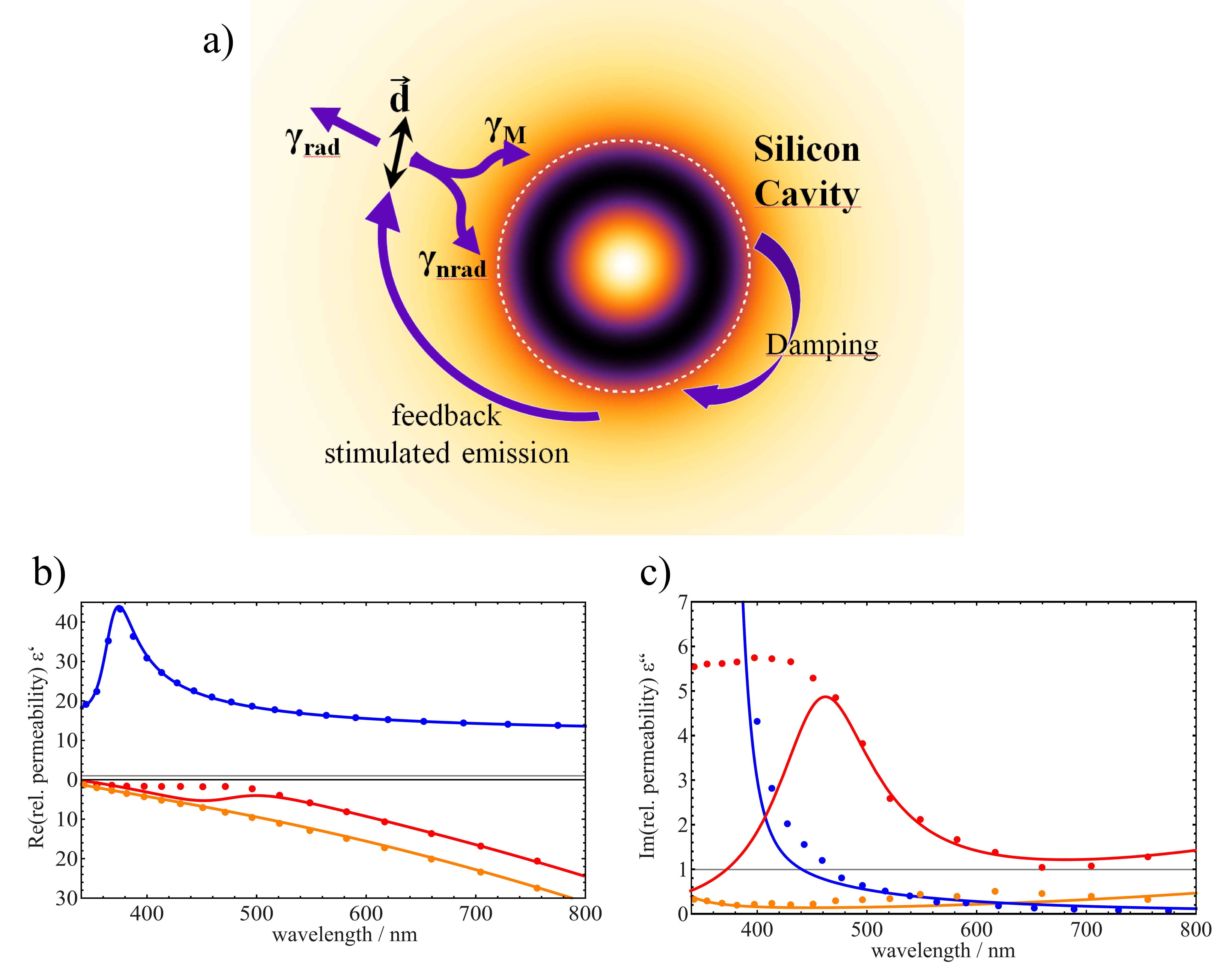}
\caption{\label{fig1}Physical model of gain medium and loss processes for a nanolaser/spaser that operates between emitters and a Mie-resonance of a dielectric/metallic sphere. 
(a) The gain-medium model considers dipolar emitters ($\vec{d}$) in proximity to a lossy resonator, specifically the decay rates into different channels like off-resonant higher order modes ($\gamma_{nrad}$), farfield ($\gamma_{rad}$) and into the resonator-mode ($\gamma_{M}$). 
(b) Real and (c) imaginary part of the dielectric permittivity of gold (red), silver (orange) and silicon (blue). Dots show experimental data \cite{Johnson1972b, Aspnes1983}, solid lines show the material models used in this study.}
\end{figure}

Our model, briefly reviewed here, is a rate equation model for the photon number $N$ and the inversion $D_E$:
\begin{equation}
\frac{{d{D_E}}}{{dt}} = {N_E}({\gamma _p} - {\bar \gamma _t}) - {D_E}({\gamma _p} + {\bar \gamma _t} + 2X{\bar \gamma _M}N)\
\label{eq1}
\end{equation}
\begin{equation}
\frac{{dN}}{{dt}} = {D_E}{\bar \gamma _M}(N + {\textstyle{1 \over 2}}) - \frac{{{Re(\omega _{\rm res})}}}{{Q}}N + {\textstyle{1 \over 2}}{\bar \gamma _M}{N_E}
\label{eq2}
\end{equation}
$N_E$ denotes the total number of two-level dipolar emitters that represent the gain medium here. These emitters are assumed to be randomly oriented and homogeneously distributed over a thin shell around the spherical particles. This realistic assumption enables us to derive an effective single-mode rate equation model (eq. \ref{eq1} and \ref{eq2}) that contains averaged decay rates $\bar \gamma_t$ (total decay especially capturing lossy decay channels) and $\bar \gamma_M$ (actual decay into the active laser mode) for the whole ensemble of emitters. $\gamma_p$ denotes the overall excitation or pump rate assuming an equal excitation for each emitter for simplicity. 
The parameters $\omega_{\rm res}$ and $Q$ are the complex eigenfrequency and quality factor of the laser mode under investigation, respectively. $X$ denotes the energy degeneracy of that very eigenfrequency, the $n$th eigenfrequency will have a $X(n)$=2$n$+1 fold degeneracy (dipole: $X$=3; quadrupole $X$=5). In our calculations we assume only a single resonance with distinct $Q$ to give rise to a feedback (stimulated emission processes), whereas there are resonances (with lower $Q$, higher degeneracy $X$ and poorer spectral overlap to the emitters) that very well lead to additional spontaneous decay. A sketch of our model is depicted in figure \ref{fig1}a). 

To begin, we examine the dielectric properties of gold, silver and silicon (figure \ref{fig1}b,c)). First, this is to show the range of validity of our material models (Drude-Lorentz models for metals and modified Lorentz model for silicon \cite{Deinega2012}) that are used in this study. Secondly and mainly, we will demonstrate that everything what follows can be estimated by carefully considering such material properties. Please note, that we refer to best scenario metals as measured by Johnson \& Christy \cite{Johnson1972b}. For silicon we use data measured by Aspnes \cite{Aspnes1983} whereas others found even smaller extinction coefficients \cite{Vuye1993}. This underlines that the following results represent a conservative estimate. From figure \ref{fig1}b) it can be seen, that the real part of silicon's relative permittivity $\epsilon'$ and thus its refractive index $n_r$ is rather large for frequencies in the optical regime, roughly $n_r$=4, meaning that the wavelength of light is four times shorter in this medium so that localized modes will also be of smaller extent. Furthermore the imaginary part $\epsilon''$ is surprisingly small even though silicons bandgap is only \SI{1.1}{eV} wide. This is due to the indirect bandgap of silicon and explains the observation of relatively high-quality resonances at photon energies above \SI{1.1}{eV}. The imaginary part of the relative permittivity $\epsilon''$ ($\epsilon''$=2$n_r\kappa$) is the relevant number when silver and gold are compared to silicon. Silicon exhibits an extinction coefficient $\kappa$ that is orders of magnitude lower but electromagnetic fields will more efficiently penetrate the dielectric material due to its high refractive index leading to higher overall losses as it would be inferred by  $\kappa$ alone. However, in terms of $\epsilon''$ silicon is still as good as the best plasmonic metal silver (except for the ultraviolet (UV) part of the spectrum) and consequently better than gold. 

\begin{figure}[h]
 \centering
\includegraphics[width=0.85\textwidth]{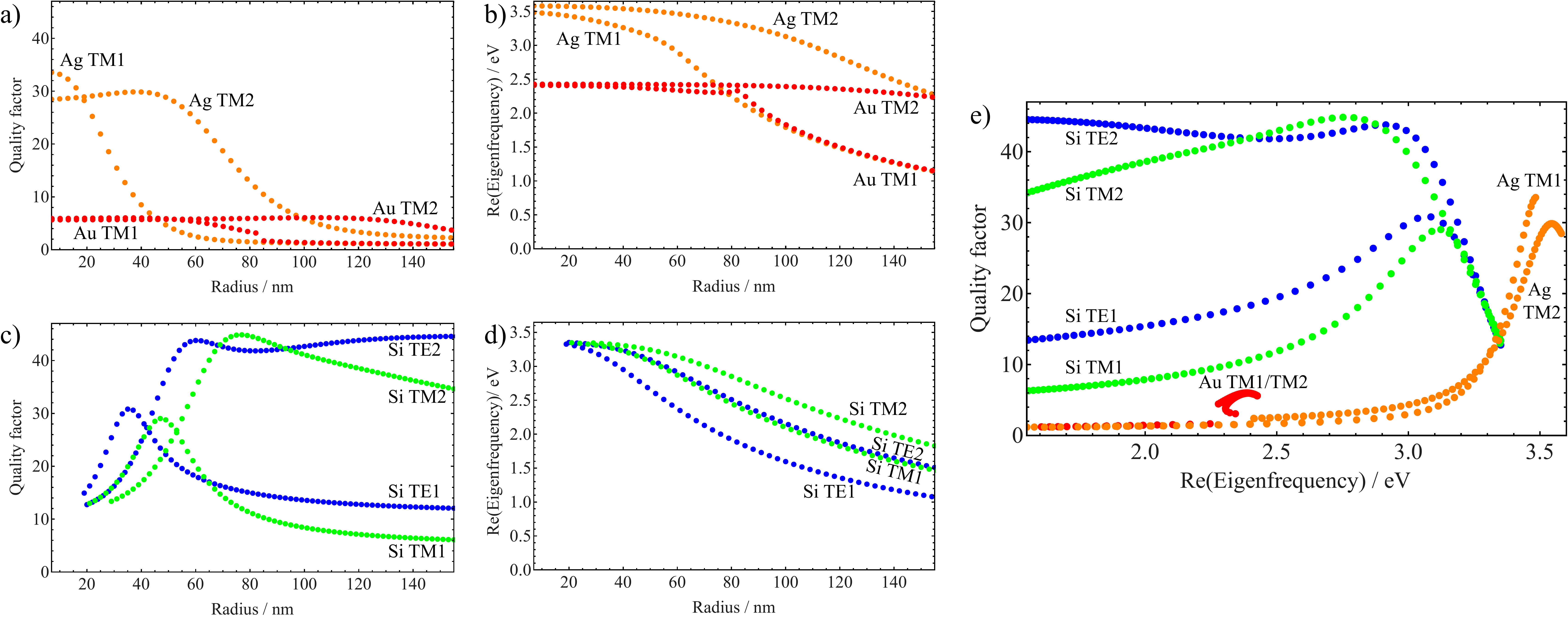}
\caption{\label{fig2}Eigenfrequencies and $Q$-factors of spherical nanoresonators made form gold (red), silver (orange) and silicon (blue, green) in vacuum. 
(a) $Q$-factors and (b) eigenfrequencies of dipolar modes (TM1) and quadrupolar modes (TM2) of gold (Au) and silver (Ag) as a function of sphere radius. 
(c) $Q$-factors and (d) eigenfrequencies of electric and magnetic dipolar modes (TM1, TE1) and quadrupolar modes (TM2, TE2) of silicon (Si) as a function of sphere radius. 
(e) shows the $Q$-factors of all three materials as a function of eigenfrequency. Silicon nanospheres support Mie modes of relatively high $Q$-factors over the entire visible spectral range. }
\end{figure}

In figure \ref{fig2}, we show how these material properties affect the actual eigenfrequencies and $Q$-factors of a sphere as a function of its $R$. Eigenfrequencies are computed by finding the roots of the sphere's scattering matrix in the complex frequency plane \cite{Kolwas2006}, the $Q$-factors are given by $Re(\omega_{\rm res})/|2Im(\omega_{\rm res})|$. We investigate the dipolar modes TE1 and TM1 and the dark quadrupolar Mie-modes TE2 and TM2 of silicon. As no TE Mie-modes exist in plasmonic nanoparticles, we compare this to the two lowest-order TM Mie-modes for each of the considered metals, silver and gold, respectively. Apart from the the UV range (roughly beyond \SI{3.2}{eV}) silicon nanospheres exhibit the largest $Q$-factors (figure \ref{fig2}a,c). Surprisingly, silicon nanospheres are not always larger than their metallic counterparts when looking at a fixed frequency and mode (figure \ref{fig2}b,d). Furthermore, while silicon spheres do show significantly improved $Q$-factors for dark modes, metal spheres do not, in contrast to what is often claimed. The reduction in radiative losses of the cavity are compensated by increased ohmic losses due to the stronger confinment experienced by the EM field in a metal. A second interesting feature becomes clearly visible, when plotting the $Q$-factors as a function of their eigenfrequencies. Silicon nanosphere resonances span the entire visible range while at the same time \textit{preserving} their high    $Q$-factors (figure \ref{fig2}e). In contrast, metallic nanospheres with reasonable $Q$-factors are limited to a rather tight spectral range. It is well known that nanorods exhibit larger $Q$-factors than spheres (up to approximately 20 for gold) also at longer wavelengths \cite{Sonnichsen2002}. However, little is known about such beneficial effects on the $Q$-factors of dark modes. Also it is beyond the scope of this paper to investigate different shapes of silicon nanoparticles to find better $Q$-factors. In any case, silicon particles can reach high $Q$-factors over the entire visible range already with the simplest design, a sphere. 

\begin{figure}[h]
 \centering
\includegraphics[width=0.85\textwidth]{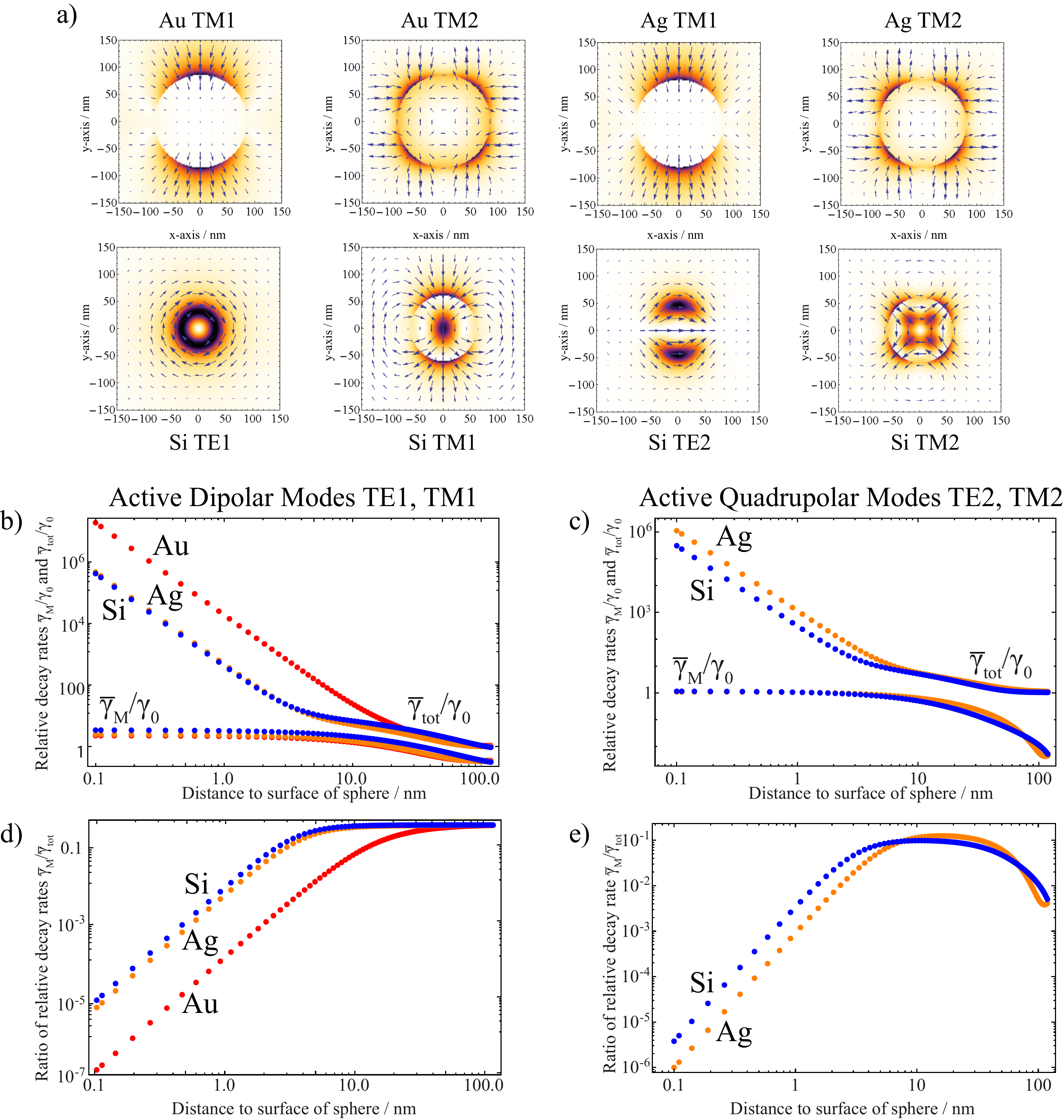}
\caption{\label{fig3}Field distributions of considered modes and relative decay rates (normalized to the decay rate in vacuum $\gamma_0$) of emitters radiating in close vicinity to a resonant spherical cavity (in vacuum) as a function of distance (convergence of rates was found after summing the first 4000 Mie-coefficients).
(a) Field intensity distributions: the first two modes of a gold and silver sphere, respectively and of the first four modes of a silicon sphere. Arrows are indicating a snap-shot of the electric vector field. 
(b) The radii and $Q$-factors are $R_{Au}=\SI{85}{nm}$ and $Q_{Au}$=2.48 for gold (red), $R_{Ag}=\SI{80}{nm}$ and $Q_{Ag}$=1.49 for silver (orange), and $R_{Si}=\SI{65}{nm}$ with $Q_{Si}$=17.02 for silicon (blue), respectively. All spheres have the same resonance frequency of \SI{2.245}{eV} corresponding to \SI{550}{nm} vacuum wavelength. The upper curves show the total (angle and orientation averaged) decay rates $\bar \gamma_t$, the lower curves the averaged decay rates into the resonant cavity-modes $\bar \gamma_M$. (c) same as (b) for an emitter resonant to a quadrupolar mode (TE2 for silicon (blue), TM2 for silver (orange)) at a frequency of \SI{2.91}{eV} corresponding to \SI{425}{nm} (gold spheres don't support modes at that frequency) with  $R_{Ag}=\SI{115}{nm}$ with $Q_{Ag}$=3.73, and $R_{Si}=\SI{60}{nm}$ with $Q_{Si}$=43.78. (d) and (e) ratio of $\bar \gamma_M$/$\bar \gamma_t$ for the same values as in (b) and (c), respectively.}
\end{figure}

Having derived the main parameters defining the cavity, we investigate the interaction of the gain-medium with the cavity. Relaxation rates of the excited state of emitters close to the spheres are calculated following Ruppin \cite{Ruppin1982a}. All decay rates stemming from the differently positioned and oriented dipolar emitters that built the active medium of our model can be averaged leading to a simple form: $\bar{\gamma}(r)=\frac{1}{X}\left(\frac{1}{3} \gamma_{\perp}(r)+\frac{2}{3}  \gamma_{\parallel}(r)\right)$, with $\gamma_{\perp}$ and $\gamma_{\parallel}$ denoting dipole orientations perpendicular or parallel to the sphere's surface, respectively. Figure \ref{fig3} shows these averaged decay rates into the laser/spaser mode, the total decay rates and the ratio of both for silver, gold and silicon spheres at the same resonance frequency of \SI{2.245}{eV} corresponding to \SI{550}{nm} vacuum wavelength for dipolar modes and of \SI{2.91}{eV} corresponding to \SI{425}{nm} vacuum wavelength for quadrupolar modes. For these cases, silicon exhibits the most favorable ratios of decay rates due to the lower total decay rate. While silver introduces relatively low non-radiative losses as expected and is almost as good as silicon, gold suffers from significant loss-channels already for emitter-sphere distances of more than \SI{10}{nm}. For the resonance chosen for the quadrupolar mode we even cannot find an eigenfrequency for gold particles due to gold's interband transistion in the green spectral range. Most surprisingly, the actual coupling to the cavity $\gamma_M$ is almost the same for all resonator materials. While it is often argued that plasmon resonances develop the most intense near-fields with the corresponding efficient coupling to emitters, for silicon spheres this is counterbalanced by higher $Q$-factors (increasing the Purcell factor) even though the field distribution is mainly located inside the sphere. 
Besides these aspects which -- as we shall elaborate below -- make silicon nanospheres very attractive for nanolaser applications on principal grounds, we would like to also discuss a number of practical aspects related to silicon nanocavities. Clearly, silicon is one of the best studied materials and can be fabricated with highest quality using well established techniques. For instance, a thin emitter-free spacing layer between gain-medium and cavity, which will significantly lower the threshold of the nanolaser \cite{Kewes2014a}, can very easily be implemented as silicondioxid films of around \SI{1}{nm} and thicker, grown in a well-controlled manner via thermal oxidation \cite{Yoshio2007}. Further, spherical silicon nanoparticles can straightforwardly be synthesized, e.g., from the gas phase -- however, not yet with mono-disperse size distributions as would be required for direct application in nanolasers. Therefore, appropriate filter-techniques have to be applied to obtain silicon nanoparticles with sufficiently narrow size distributions. Alternatively, the fabrication of appropriate nanostructures via e-beam lithography on silicon-on-insulator chips is straightforward and has already been demonstrated \cite{Coenen2013}. 

\begin{figure}[h]
 \centering
\includegraphics[width=0.85\textwidth]{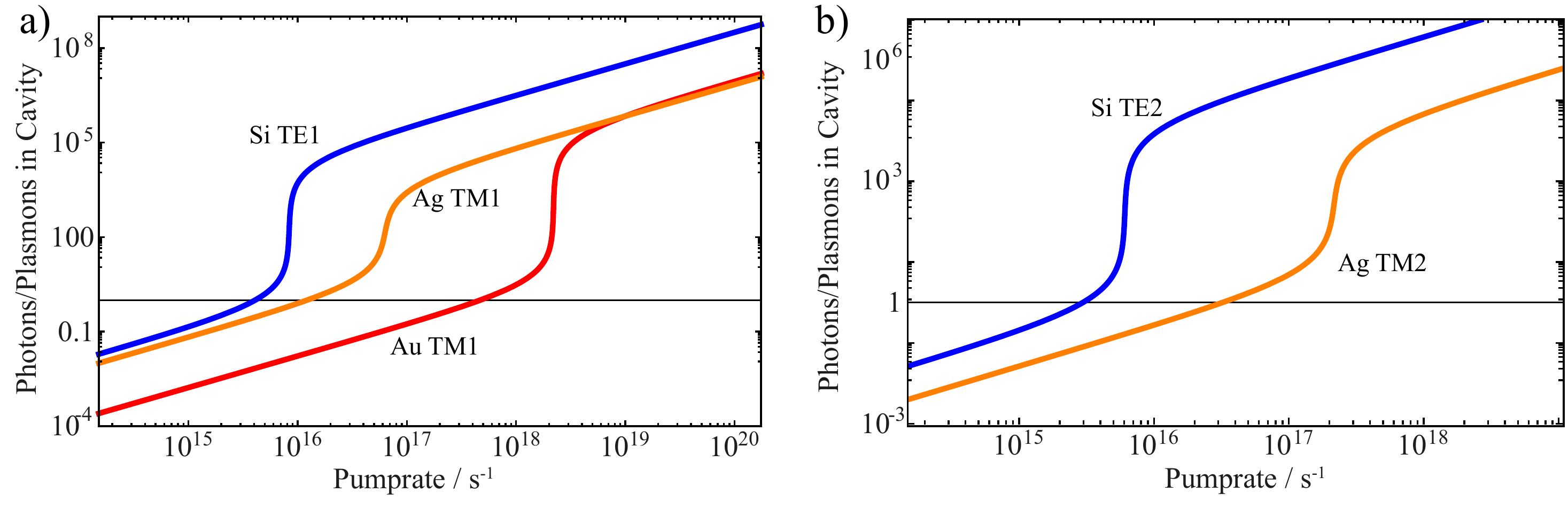}
\caption{\label{fig4}Stationary solution of laser rate equations: input-output curves $N(\gamma_p)$ for gold (red), silver (orange) and silicon (blue) cavities with the same properties as in figure \ref{fig3} coated with \SI{1}{nm} of 2000 homogeneously distributed and randomly oriented dipolar emitters for (a) the bright dipolar cavity-mode and (b) the dark quadrupolar cavity-mode.}
\end{figure}

So far, we have found that $Q$-factors are best for silicon nanospheres, that unwanted non-radiative decay-rates close to the sphere are relatively small, especially smaller than for the metall spheres and that the decay rate into the cavity is of comparable magnitude. As alluded to above, this already implies that the silicon nanolaser exhibits excellent performance characteristics, also when compared to spasers. 
The analytic solutions for the stationary problem (d$N$/d$t$=d$D_E$/d$t$=0), i.e., input-output curves (photon number in the cavity as a function of pump-rate $N(\gamma_p)$) of the rate equations and the sphere radii and frequencies used in the above calculations of $Q$-factors and decay rates are depicted in figure \ref{fig4}. A \SI{1}{nm} layer of gain medium is surrounding the spheres. All lasers/spasers have the same absolute number $N_E$ of 2000 emitters (slightly different emitter densities due to the different radii in the realistic range of \SI{0.02}{nm^{-3}} and \SI{0.04}{nm^{-3}} for layer-by-layer approaches). Both, for the dipolar and the quadrupolar cavity mode, we find indeed the lowest threshold for the silicon nanolaser. 

\begin{figure}[h]
 \centering
\includegraphics[width=0.85\textwidth]{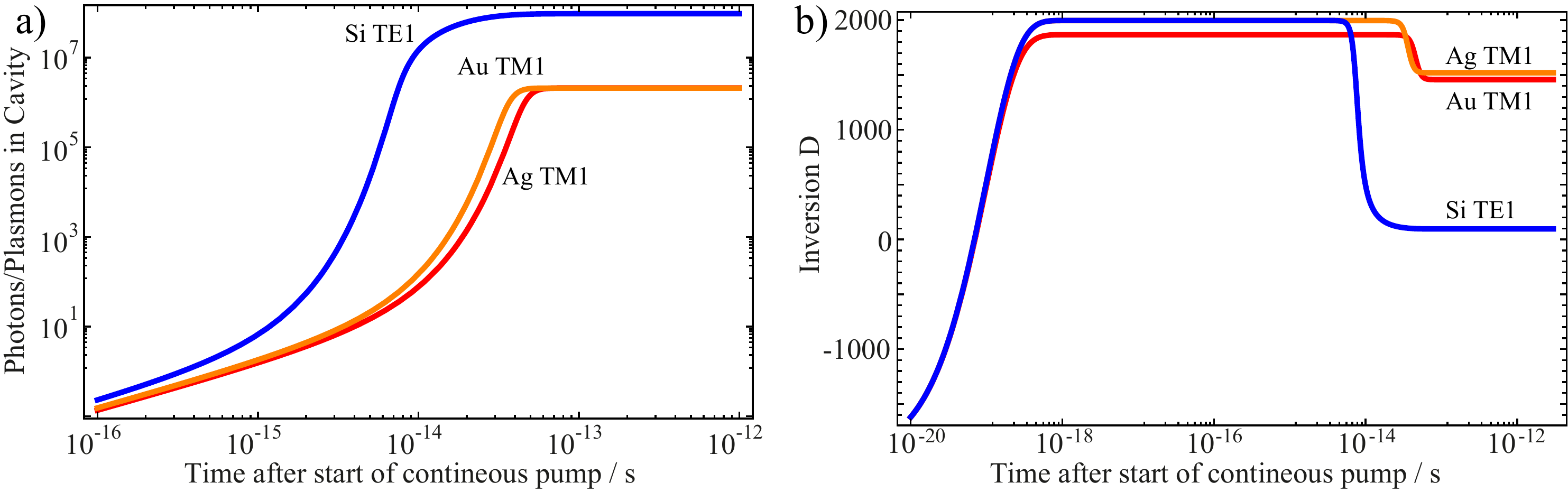}
\caption{\label{fig5}Time-dependent solution of laser rate equations for the bright dipolar cavity-mode: (a) input-output curves $N(\gamma_p)$ for gold (red), silver (orange) and silicon (blue) cavities with the same properties as in figure \ref{fig3} coated with \SI{1}{nm} of 2000 homogeneously distributed and randomly oriented dipolar emitters. (b) Temporal evolution of the inversion.}
\end{figure}

It has been discussed by Stockman \cite{Stockman2010}, that the spaser could behave as an ultra fast all optical switching device as the times to reach the threshold and to return to normal fluorescence can be extremely short. By solving the dynamic rate equations numerically, we also can investigate this effect within our model to give a qualitative comparision of spasers and lasers. For the solution of the time-dependent rate equations we choose the starting values at $t_0$ of no photons/plasmons in the resonator and no emitters in the excitated state ($N(t_0)$=0 and $D_E(t_0)$=-$N_E$). With these starting values we simulate a relaxed system that suddenly gets excited with a continuous pump. Figure \ref{fig5} shows the time-dependent evolution of both the photon/plasmon number and the inversion for the lasers/spasers working resonant to the dipolar modes investigated before. Again the silicon based nanolaser can compete with the spasers, in fact the switch-on time is even faster. 

Compared to nanowire lasers, the silicon nanolaser is obviously smaller. Apart from that, the suggested gain medium will presumably differ. The resonator of a nanowire lasers is already the gain medium, thus a good compromise between material quality, high refractive index and bandgap has to be found to achieve a small, low threshold laser at a designated wavelength. Even then, the amount of gain that has to be pumped at least is fixed by the geometry. In contrast to that, our proposal represents a hybrid device: the silicon nanoresonator is a passive cavity that will be presumably combined with a well defined amount of organic gain that typically provides strong oscillator strength. Thus, due to the broad-band availabilty of such organic gain media, the laser is tunable over the entire visible range and can be potentially operated with low energy consumption. 

While plasmonic nanoparticles only support modes of electric type (TM) modes, silicon nanoparticles support both, electric and magnetic (TE) modes, rendering silicon nanolasers a complementary nanoscopic coherent light source to the spaser. Silicon nanolasers working resonant to TE modes might be used as loss compensating elements in metamaterials, providing at the same time a magnetic response comparable to split-ring resonators \cite{Kuznetsov2012}. Furthermore, exploiting the recently demonstrated \cite{Person2013} Kerker condition \cite{Kerker1983}, which is based on simultaneously having access to both, TE and TM modes, might enable simple and compact lasing structures that exhibit a directional emission. 

In conclusion, we have shown that a silicon nanoparticle based nanolaser represents an attractive candidate for a coherent light-source that is nanoscopic in all three dimensions. Due to the few intrinsic losses and its high refractive index, silicon nanoparticles can support electromagnetic Mie-modes of relatively high $Q$-factors over the entire visible spectral range. Because of this, relatively high electromagnetic field intensities are retained around the nanoparticle, mandatory to achieve sufficient coupling of the gain-medium to the cavity. The low losses of silicon furthermore inhibits excessive coupling to lossy decay channels that can significantly increase the threshold (as would be the case for spasers \cite{Kewes2014a}). Furthermore we have investigated lasing in bright and dark modes as well as switch-on times of the lasing systems. In all cases, we have found that silicon nanolasers exhibit surprisingly good performance characteristics and thus can compete with spasers in almost all aspects, potentially even outperforming them in certain aspects. 
 
\acknowledgement
We acknowledge support by the Deutsche Forschungsgemeinschaft (DFG) through the subprojects B2 and B10 within the Collaborative Research Center (CRC) 951 Hybrid Inorganic/Organic Systems for Opto-Electronics (HIOS). 

\bibliography{si-nanolaser_arxiv}

\end{document}